\documentclass[aps,prd,superscriptaddress,nofootinbib,floats,amsmath,showpacs,showkeys]{revtex4}

\usepackage{psfrag}
\usepackage{graphicx}
\usepackage{hyperref}

\begin{document}

\title{Primordial black hole production during preheating in a chaotic inflationary model}

\author{E. Torres-Lomas }
\email[]{efrain@fisica.ugto.mx}

\author{L. Arturo Ure\~na-L\'opez}
\email[]{lurena@fisica.ugto.mx}
\affiliation{Departamento de F\'isica, DCI, Campus Le\'on, CP 37150,
Universidad de Guanajuato, Le\'on, Guanajuato, M\'exico}


\begin{abstract}
In this paper we review the production of primordial black holes
(PBHs) during preheating after a chaotic inflationary model. All
relevant equations of motion are solved numerically in a modified
version of HLattice, and we then calculate the mass variance to
determine structure formation during preheating. It is found that
production of PBHs can be a generic result of the model, even
though the results seem to be sensitive to the values of the smoothing
scale. We consider a constraint for overproduction of PBHs that could
uncover some stress between inflation-preheating models and observations. 
\end{abstract}

\maketitle

\section{Introduction}
Recent research about the implications of a preheating stage on
observable cosmological parameters (CMB, microhalos of dark matter,
non-gaussianity) have been reported. In particular PBHs
formation has been analysed in different models, it was found that an
overproduction of these objects may appear but no conclusive results
have been reported
\cite{Erickcek:2011us,Finelli:1998bu,Jedamzik:2010hq}.

The primordial black holes (PBHs) production during inflation can play an important role in the
physics after inflation, for example, by producing the reheating of the
universe \cite{Hidalgo:2011fj}. This PBHs production may appear only in 
some inflationary theories but not in a generic model. Despite of this,
PBHs formation can occur during a preheating phase at the graceful exit
from inflation \cite{Bassett:1998wg, Bassett:1999mt, Green:2000he, Khlopov:1985jw,
  Bassett:2000ha, Suyama:2006sr}. The PBHs production that appears due
to the explosive excitement of matter scalar perturbations has been
widely explored both analytically and numerically, in cases in which
metric perturbations are not taken into account. On the other hand,
the production of the metric perturbations also has been studied but
for different reasons, but with minor progress in the numerical part.

The complete analysis of the dynamics of a Universe during the
preheating stage, with the consideration of both the redispersion and the feedback with the 
perturbations of the metric, can only be addressed by numerically solving the full Einstein's equations. In this
paper we show the results of numerical evolution of the mass variance
as a measure of structure formation in a model with two scalar fields,
an inflaton scalar field $\phi$ and a spectator $\chi$, where the
potential for chaotic inflation $\sim m^2 \phi^2$ and an interaction
term $\sim \phi^2 \chi^2$  have been considered. The numerical
solution is calculated by HLattice code \cite{Huang:2011gf}, which
realizes a full 3dimensional integration of the equations of
motion. In this respect, our results do not rely on the assumption
that matter and metric perturbations evolve separately.

A description of the paper is as follows. First of all, we provide the
mathematical basis for the general study of preheating
models. Throughout the numerical simulations, matter and metric
perturbations are calculated, and the particle production associated
with each scalar field at the end of the simulation is also
reported. The initial and final spectrum of the curvature perturbation
and the evolution of the mass variance are calculated as a measure of
structure formation, with special interest in the conditions under
which we may have  PBHs overproduction: $\sigma \ge\sigma_*$. Finally,
we discuss the cosmological implications that may arise from such an
overproduction of PBHs.

\section{Preheating \label{sec:preheating-}}
We consider a preheating model where the matter Lagrangian ${\cal
  L}_{mat}$, that includes an inflaton field $\phi$, and a spectator field
$\chi$, is given by
\footnote{The name \emph{spectator} comes from considering that during
  inflation the whole dynamic was determined solely by the inflaton
  field $\phi$. However, a more realistic treatment would also
  consider the effects of the spectator $\chi$ field during
  inflation.}
\begin{equation}\label{eq:1}
  {\cal L}_{mat}= \frac{1}{2} \phi_{,\alpha} \phi^{,\alpha}  +
  \frac{1}{2} \chi_{,\alpha}\chi^{,\alpha}  + \frac{1}{2} m^2\phi^2 +
  \frac{1}{2} g^2 \phi^2 \chi^2 \, , 
\end{equation}
where $m=10^{-6}M_{Pl}$, which is an typical mass value for the
chaotic inflationary model \cite{Bennett:2012fp}, and $g^2$ is
the coupling constant between the scalar fields. We also assume a flat
background Friedmann-Robertson-Walker (FRW) geometry with metric
perturbations given in the synchronous gauge as \cite{Ma:1995ey}
\begin{equation}
  ds^2 = a^2(\tau)\left(-d\tau^2 + (\delta_{ij} + h_{ij})dx^i
    dx^j\right) \, , \label{eq:2}
\end{equation}
where the metric perturbation $h_{ij}$ can be decomposed as
\begin{equation}
  h_{ij}=h\delta_{ij}/3 + h^\parallel_{ij} + h^\perp_{ij} + h^T_{ij}
  \, , \label{eq:3}
\end{equation}
where $h\delta_{ij}/3 + h^\parallel_{ij}$ represents the scalar part
of the perturbation, whereas $h^\perp_{ij}$ and $h^T_{ij}$ represent
the vectorial and tensorial parts, respectively.

In this work, our main concern are scalar perturbations, so it will be
useful to define the parametrization of the scalar part
$h_{ij}^{ \left[ S \right] } \equiv h\delta_{ij}/3 + h^\parallel_{ij}$ in Fourier
space as
\begin{equation}
  h_{ij}^{ \left[ S \right] } (\vec{x},\tau) = \int d^3k e^{i\vec{k}\cdot\vec{x}}
  \left\{ \hat{k}_i\hat{k}_j h(\vec{k},\tau) +
    (\hat{k}_i\hat{k}_j - {1 \over 3}\delta_{ij})\,
    6\eta(\vec{k},\tau) \right\} \, , \quad \vec{k} =
  k\hat{k} \label{eq:4} 
\end{equation}
where the trace $h$ is used for both the real space and the Fourier
space. It is convenient to define also the curvature perturbation in
Fourier space, $\zeta_k$, as 
\begin{equation}\label{eq:zeta}
  \zeta_k \equiv -\eta_k +  {\delta \rho^S_k \over 3 (\langle \rho \rangle + \langle p \rangle)} \, , 
\end{equation}
which by definition is a gauge-invariant quantity \cite{PhysRevD.28.679, PhysRevD.62.043527}.
Here, $\langle \rho \rangle$ and $\langle p \rangle$ are the mean values
of the energy density and the pressure, respectively, that are defined as
\begin{eqnarray}
\langle \rho \rangle & = & \langle \frac{1}{2}\dot{\phi}^2+\frac{1}{2}\dot{\chi}^2+\frac{1}{2a^2}|\nabla \phi|^2 
+ \frac{1}{2a^2}|\nabla \chi|^2 +\frac{1}{2}m^2\phi^2 + \frac{1}{2}g^2\phi^2 \chi^2 \rangle ,\\
\langle p \rangle & = & \langle \frac{1}{2}\dot{\phi}^2+\frac{1}{2}\dot{\chi}^2-\frac{1}{6a^2}|\nabla \phi|^2 
- \frac{1}{6a^2}|\nabla \chi|^2 -\frac{1}{2}m^2\phi^2- \frac{1}{2}g^2\phi^2 \chi^2 \rangle ,
\end{eqnarray}
where $\langle \cdot \rangle$ means the spatial averaging in the 3
dimensional lattice.

We have calculated numerically the time evolution of the curvature
perturbation (see Eq.~(\ref{eq:zeta})) for a Universe in the epoch of
preheating, using the free numerical code HLattice
\cite{Huang:2011gf}. This code realizes a full integration of the
Einstein-Klein-Gordon equations of motion for two scalar fields,
within a 3+1 set up in the synchronous gauge; a complete analysis of
the numerical methods used and the accuracy of the numerical results
can be found in\cite{Huang:2011gf}. The metric perturbations are
evolved by solving the Einstein equations of this model, without any
explicit separation in homogeneous and inhomogeneous variables. 

The initial conditions are such that the inflaton is oscillating around the
minimum of the potential. The initial fields are initialized with a random 
gaussian fluctuations according to values a Bunch-Davis vacuum
\cite{Bunch21031978}.  The value of the mass and the coupling constant
are $m = 10^{-6} M_{pl}$ and $g^2 = 2.5 \times 10^{-7}$,respectively.  

\begin{figure}
\centering 
 \includegraphics[height=5.5cm, width=0.49\textwidth]{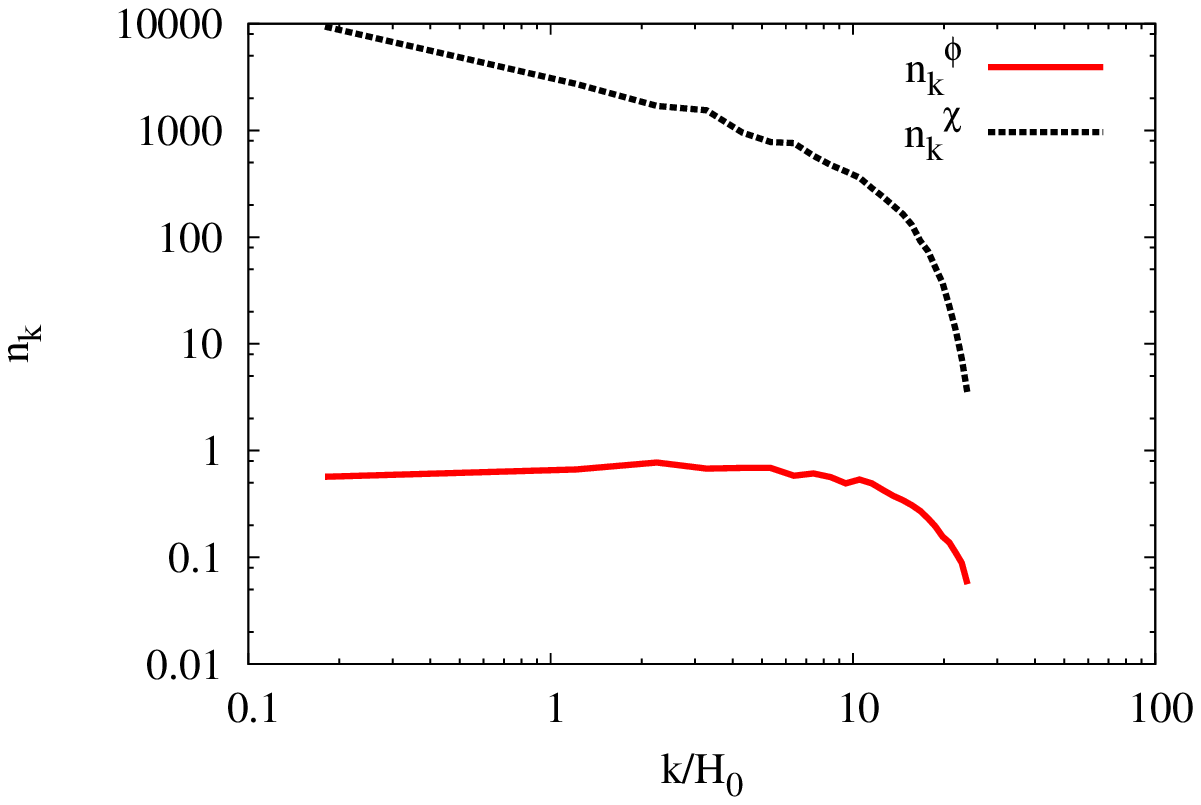}
 \includegraphics[height=5.5cm, width=0.49\textwidth]{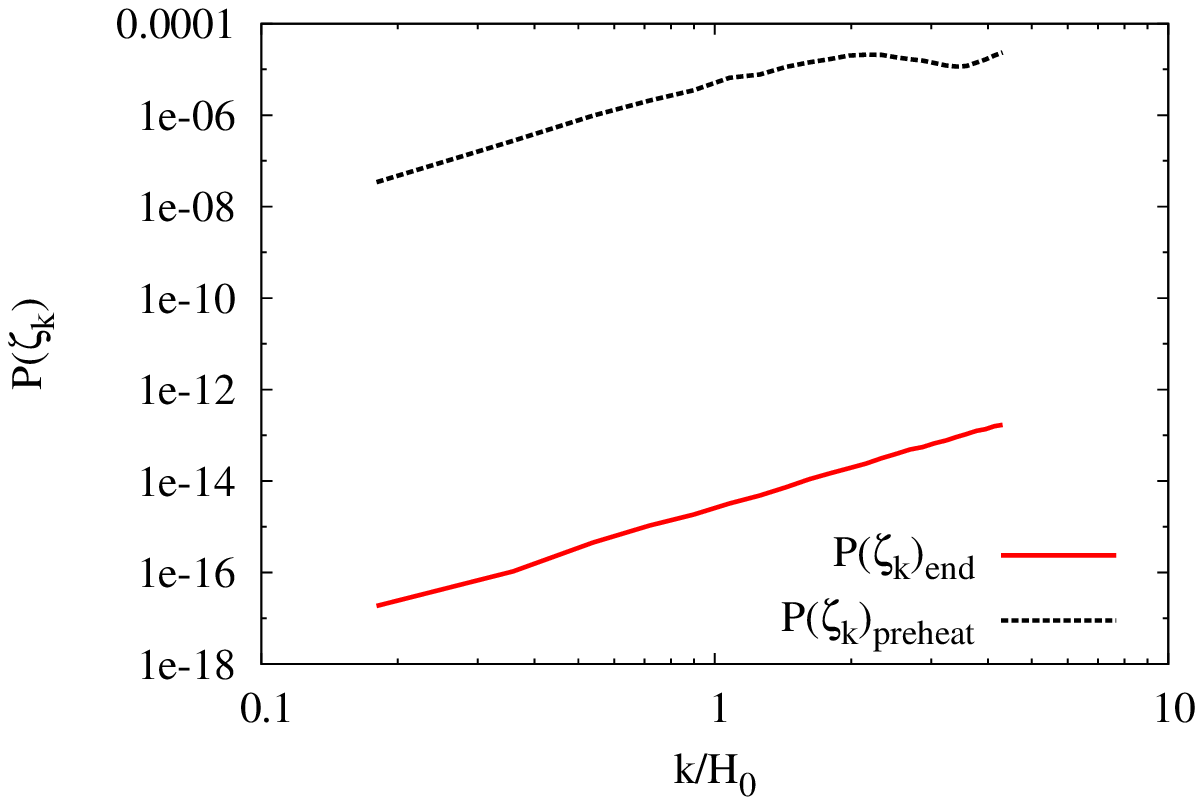}
\caption{(Left) Final spectrum of occupation numbers $n_k^\phi$ and $n_k^\chi$. 
This growth of $n_k^\chi$ is produced by parametric resonances. The production 
of light particles with a high kinetic energy can increase the
temperature to produce a radiation dominated Universe.  
(Right) The power spectrum of curvature perturbation at the end
of inflation $P(\zeta_k)_{end}$ and at the end of preheating
$P(\zeta_k)_{reh}$. This growth of the curvature perturbation should
produce PBHs formation. See the text for more details.}
\label{graph:occupation_curvature}
\end{figure}

To exhibit the dynamics of the matter during preheating
Fig.~\ref{graph:occupation_curvature} (left) shows the final spectrum
of occupation numbers $n_k^\phi$ and $n_k^\chi$ associated with the
scalar fields $\phi$ and $\chi$, respectively. This growth of
$n_k^\chi$ produced by parametric resonances has been previously
reported in the literature \cite{Dufaux:2006ee, Podolsky:2005bw}. In
Fig~\ref{graph:occupation_curvature} (right) the power spectrum of
curvature perturbation at the end of inflation $P(\zeta_k)_{end}$ and
at the end of preheating $P(\zeta_k)_{reh}$ are shown. The growth of
$P(\zeta_k)$ is a product of the growth of both the matter and metric
perturbations (See (\ref{eq:zeta})). This growth of the curvature
perturbation  should produce the structure formation that could
possibly evolve to form PBHs.

\section{Primordial black holes  formation \label{sec:pbh-formation-}}

If density perturbations acquire sufficiently large values, they can
cause collapsing regions to produce early structure, causal isolated regions,
and possibly primordial black holes
\cite{Hawking:1971ei,Zeldovich:1967,Carr:1975qj}.

The problem of finding the value of the critical overdensity
$\delta_c$ in perturbations such that a primordial black hole can be
formed in the early Universe has been discussed by several authors
both analytically and numerically\cite{Niemeyer:1999ak,
  Shibata:1999zs, PhysRevLett.80.5481, Musco:2004ak,
  Green:2004wb}. The progress made in this regard is vague and
sometimes contradictory. For example, in 1975 Carr \cite{Carr:1975qj}
obtained that $\delta_c \sim w$, where $w$ is the equation of state,
while in \cite{Niemeyer:1999ak,Shibata:1999zs} Niemeyer \& Shibata
obtained that the critical value is $\delta_c>0.7$ and in
\cite{Green:2004wb} Green conclude that a smaller value, $\delta_c
\sim 0.3$, is a better estimation, this last coincides with the Carr's
results for the case of a Universe in radiation domination epoch.

On the other hand, to estimate the mass of a primordial black holes is
necessary to calculate the black hole mass function $\beta$ defined as
\begin{equation}\label{mass_function}
  \beta(M) = \frac{\rho_{PBH}}{\rho_{Tot}} = \int_{\delta_c}^{\infty}
  P(\delta) d\delta  \, ,
\end{equation}
which measures the contribution of PBH to the total energy density of
the Universe. This parameter is constrained by observational evidence
over a wide range of masses; the stronger constraint lies on PBHs of
mass $M \sim 10^{15}g$, whose evaporation time is $\tau \sim$ age of
the Universe, so $\beta \leq 10^{-20} $, a value that basically comes
from the non-observation of $\gamma$-rays originating on Hawking
radiation \cite{Hawking:1971ei,Carr:2005zd}.

Following typical calculations, if the over-dense regions are
spherically symmetric with a Gaussian distribution given by
\begin{equation}
  P(\delta) = {1 \over \sqrt{2 \pi} \sigma} \exp{\left( -{\delta^2
        \over 2\sigma^2} \right)} \, , \label{eq:6}
\end{equation}
where $\sigma$ is the mass variance at horizon crossing, it is
possible to transfer the observational constraint on PBHs from the mass
function $\beta$ to the mass variance $\sigma$. Thus, we get a
threshold mass variance of $\sigma_*=0.08$, any mass variance above
this value indicates an over-production of PBHs which would put a
tension between the inflation-preheating model and observations. If instead 
of a Gaussian distribution (see Eq.~(\ref{eq:6})) a first order
chi-squared distribution for density fluctuations is considered, the threshold
changes to $\sigma_*=0.03$ \cite{Green:2000he}.

Given the curvature perturbation (\ref{eq:zeta}), it is possible to define
the mass power spectrum by
\begin{equation}\label{spectrum_zeta}
  {\cal P}_\zeta \equiv {k^3 \over 2 \pi^2} \left| \zeta_k \right|^2
  \, , 
\end{equation}
and the mass variance $\sigma$, see Eq.~(\ref{eq:6}), as
\begin{equation}\label{sigma}
  \sigma^2 = \left( \frac{4}{9} \right) \int_0^\infty \left(
    \frac{k}{aH} \right)^4 {\cal P}_\zeta \tilde{W}(kR) \frac{dk}{k}
  \, ,
\end{equation}
where $\tilde{W} (kR) $ is the window function in Fourier space given by
\begin{equation}
  \tilde{W} (kR) \equiv \exp{(- k^2 R^2 / 2)} \, , \label{eq:8}
\end{equation}
and $R$ is the artificial smoothing scale \cite{Liddle:1993fq}. 
\begin{figure} 
\centering 
 \includegraphics[height=5.5cm, width=0.49\textwidth]{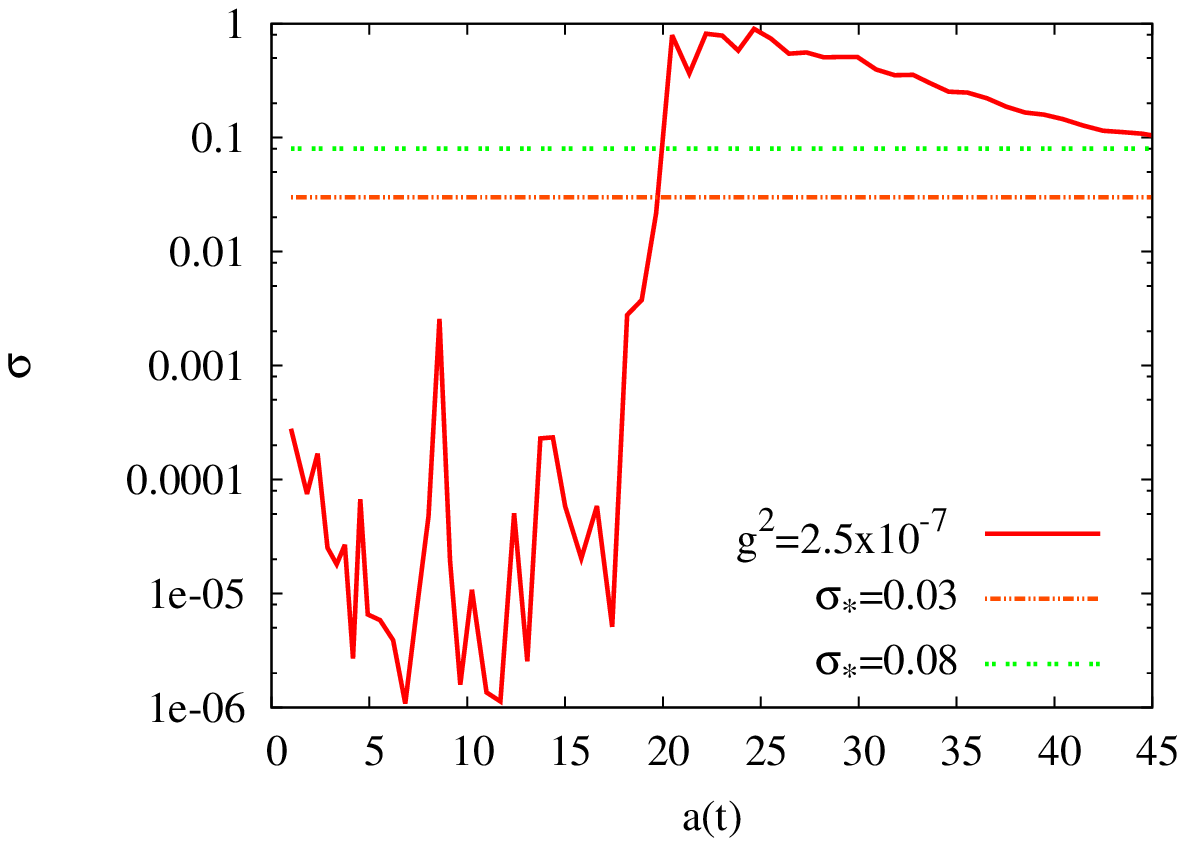}
 \includegraphics[height=5.5cm, width=0.49\textwidth]{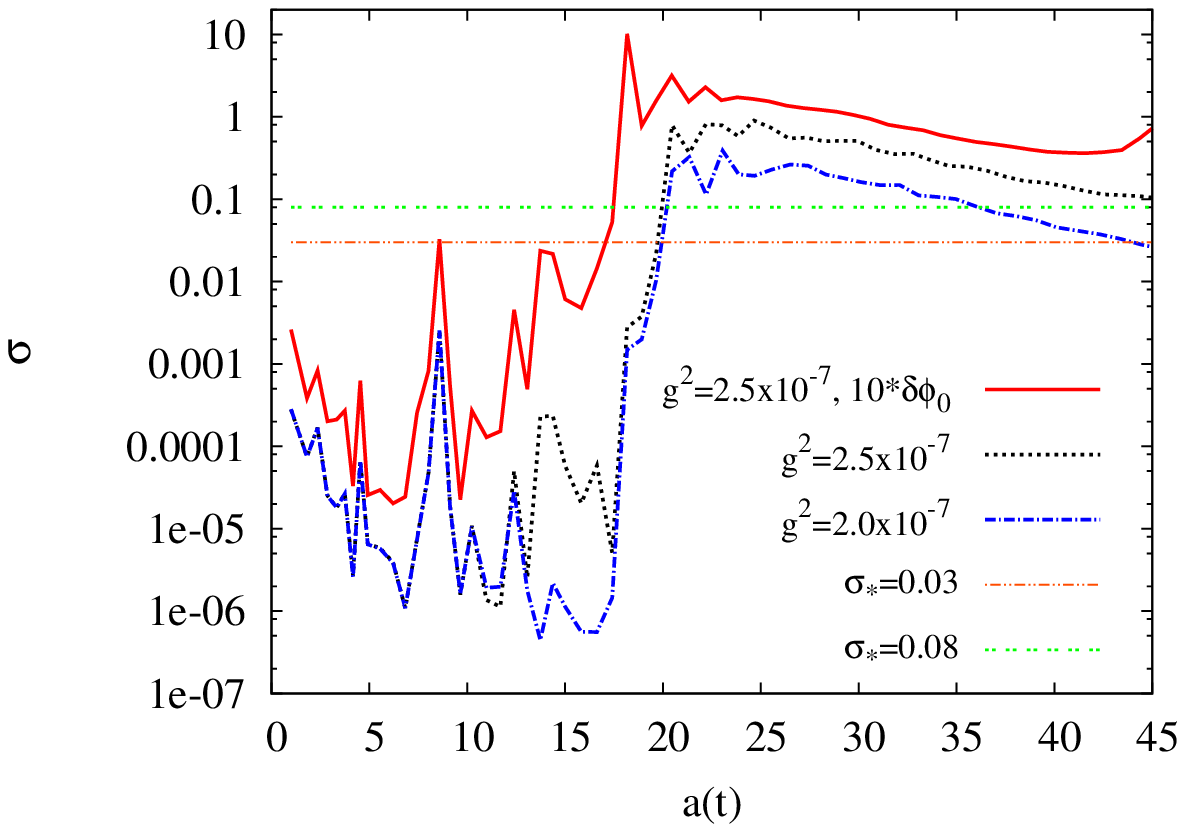}
\caption{(Left) The mass variance $\sigma$ grows explosively during the
  preheating process even exceeding both, $\sigma_* =0.03 $ and $
  \sigma_* =0.08$ values. (Right) The dependence on coupling constant
  parameter and on initial amplitude inflaton perturbations. The
  smoothing scale is $R = 1 / aH$. This suggests that this chaotic
  inflationary model presents PBHs production, although the exact
  amount depends on the chosen values of the parameters. See the
  text for more details.}
\label{graph:sigma_phi^2_model}
\end{figure}

The growth of the modes of the energy overdensity, caused by the
parametric resonances of $\delta \chi_k $ \cite{Kofman:1997yn},
produces the growth of the modes of $ \delta \rho_k$ and
$h_{ij}$. Hence, we expect a growth of the modes of $\zeta_k$, so that
the spectrum~(\ref{spectrum_zeta}) must also grow during the
preheating process. These changes should be reflected in the mass
variance, which would give us information on how much lumpy the
Universe becomes. This phenomenon was observed already
in \cite{Bassett:2000ha} for a quartic inflationary model.

Fig.~\ref{graph:sigma_phi^2_model} (left) shows the growth of $\sigma$
during the numerical evolution of the preheating model
in~(\ref{eq:1}). It is to be noted that the variance grows explosively
during the preheating process, even exceeds the aforementioned bounds
$\sigma_* =0.03$, and $ \sigma_* =0.08$, if the smoothing scale is $R
\equiv 1/k_* = 1 / aH$, see Eq.~(\ref{eq:8}). This is an evidence that
overproduction of PBHs is possible in this model of chaotic inflation
for the values of parameters that we have chosen.

The dependence of $\sigma$ on parameters of this model is shown in 
Fig.~\ref{graph:sigma_phi^2_model} (right). It is observed that the amplitude of
$\sigma$ strongly depends on both the value of the coupling constant
$g$ and of the initial field perturbations, even when the initial amplitudes 
in the initial fluctuations $10 (\delta \phi_0)$ is not a well motivated selection.\footnote{In
this case $10 (\delta \phi_0)$ means that the initial fluctuations of
the fields were tenfold increased from the Bunch-Davies vacuum.} In
this way, the requirement for the non overproduction of PBHs can be used to add
constraints on the values of the model parameters.

\begin{figure} 
\centering 
 \includegraphics[height=5.5cm, width=0.5\textwidth]{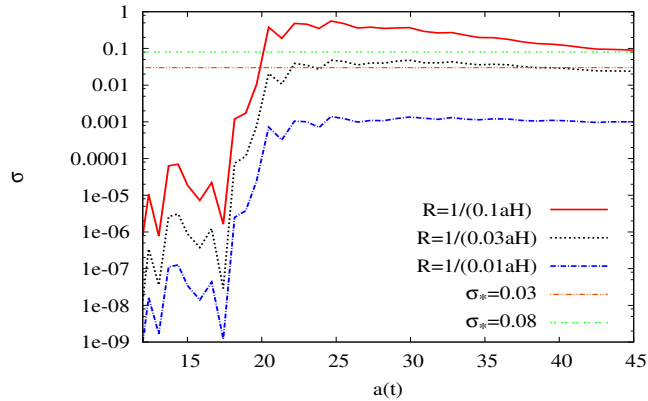}
 \caption{Time evolution of $\sigma$ for three sizes of the smoothing scale $ R $ are shown. The
  $R=1/(0.1aH)$ case shows overproduction but for $R=1/(0.03aH)$ the
  overproduction is reduced. For a artificial smoothing scale
  $R=1/(0.01aH)$ only stable structures are produced, see also the
  text for more details.}
\label{graph:R_0_01}
\end{figure}

In Fig.~\ref{graph:sigma_phi^2_model} we also observe that in all
cases $\sigma$ decays immediately after the end of preheating, which
could give us indications that, in this model, during the beginning of thermalization process, part
of the structure has been diluted. On the other hand, Fig.~\ref{graph:R_0_01} shows the
dependence of $\sigma$ on the size of the artificial smoothing scale $ R $. If we
change the size of the smoothing scale we find that:
\begin{enumerate}
\item The maximum variance is reduced significantly even preventing
  the overproduction of PBHs.

\item The mass variance does not decrease for $R=1/(0.01aH)$ at the
  end preheating.
\end{enumerate}
All of this likely indicates that we need to calculate a scale $R_*$
containing  only structures that are able of survive the
thermalization process. Only these stable structures would be present
at the beginning of radiation domination as very early structures.

\section{Conclusions \label{sec:conclusions-}}

In this work we have solved the system of Einstein equations for a
universe at the time of preheating in a chaotic inflationary
model. The dynamics of scalar perturbations of both matter and metrics
have been evaluated. Using the spectrum of the gauge invariant
variable $\zeta$ has been possible to calculate the mass variance $
\sigma $.

Considering the model~(\ref{eq:1}) with $m=10^{-6}M_{Pl}$
and $g^2=2.5\times 10^{-7}$, it was found that the mass variance $\sigma $
exceeds the $\sigma_*$ value. We have considered the constraint $ \sigma \le \sigma_*=
0.08$ which comes from the condition imposed upon the mass fraction $
\beta \equiv \rho_{PBH} / \rho_{TOT}$, which is in turn obtained from
the observation of the background in gamma rays. The result $ \sigma >
\sigma_*$ could put stress between this inflation-preheating model and
observations.

It has been found that the production during preheating can be a generic result
even though the results depends on the parameters of model, initial conditions and
on the smoothing scale selection. Because the $\delta_c$ value 
has not been determined in an preheating ambiance and then $\sigma_*$
value is not well  established, the PBHs over-production should be taken with some caution.

We have shown that after the preheating process the structures produced during preheating can be able to survive the
thermalization process. This structures possibly would be present at the beginning of Hot Big-Bang era.

More research is needed in this area, and we are working on the
production of PBH in a more general inflationary model. The results
will be reported elsewhere.

\bibliography{PBH_Biblio}

\end{document}